\documentclass[superscriptaddress,twocolumn,amssymb,aps,floatfix]{revtex4}

\usepackage{graphicx,dcolumn,bm,amsmath,color,bm,upgreek,hyperref,caption}

\renewcommand{\rm}{\textrm} 
\renewcommand{\bf}{\textbf} 

\newcommand{\eq}[1]{Eq.~(\ref{#1})}
\newcommand{\fig}[1]{Fig.~\ref{#1}}
\newcommand{\figs}[1]{Figs.~\ref{#1}}
\newcommand{\sect}[1]{Sec.~\ref{#1}}
\newcommand{\bQ}{\bf{Q}}
\newcommand{\Seq}{S_\rm{eq}}
\newcommand{\nv}{\hat{\bf{n}}}
\newcommand{\br}{\bf{r}}

\newcommand{\um}{~\upmu\rm{m}}

\newcommand{\Jm}[1]{~\rm{Jm}^{#1}}
\newcommand{\pN}{~\rm{pN}}
\newcommand{\dgC}{\ifmmode \text{°C} \else °C \fi}
\newcommand{\dg}{\ifmmode \text{°} \else ° \fi}

\begin{document}
\title{Landau-de Gennes numerical simulation of nematic liquid crystals utilizing radial basis functions}

\author{Jin-Sheng Wu}
\affiliation{Department of Physics and Chemical Physics Program, University of Colorado, Boulder, CO, USA}
\affiliation{International Institute for Sustainability with Knotted Chiral Meta Matter, Hiroshima Universit (WPI-SKCM2), Higashihiroshima, Japan}
\author{Ivan I. Smalyukh}
\affiliation{Department of Physics and Chemical Physics Program, University of Colorado, Boulder, CO, USA}
\affiliation{International Institute for Sustainability with Knotted Chiral Meta Matter, Hiroshima University, Higashihiroshima, Japan}
\affiliation{Department of Electrical, Computer, and Energy Engineering, Materials Science and Engineering Program and Soft Materials Research Center, University of Colorado, Boulder, CO, USA}
\affiliation{Renewable and Sustainable Energy Institute, National Renewable Energy Laboratory and University of Colorado, Boulder, CO 80309, USA}

\email{ivan.smalyukh@colorado.edu}

\begin{abstract}
Numerical simulations based on radial basis functions have been developed for systems with complex geometries and have been successfully applied across various fields, including seismology, coastal hydrodynamics, and biology. However, examples in liquid crystal modeling are limited. In this study, we present a Landau-de Gennes numerical simulation of nematic liquid crystals utilizing radial basis functions, emphasizing its advantages over traditional cubic grid calculations, such as enhanced geometric flexibility and improved computational efficiency. Through simulations of liquid crystal-colloid systems with diverse geometries, we demonstrate that our approach effectively captures the essential topological and energetic features of liquid crystal equilibrium structures. Additionally, we introduce an adaptive node refinement scheme that is crucial for resolving the fine structure of singular defects in nematic liquid crystals.
\end{abstract}

\maketitle

\section{Introduction}
Radial basis functions (RBFs) provide a powerful numerical methodology for solving partial differential equations via an elaborate technique that utilizes radially symmetric functions as basis functions instead of delta functions \cite{fornberg2015primer,buhmann2000radial,majdisova2017radial,gutmann2001radial}. Therefore, this approach has many advantages as compared to other numerical approaches, especially when applied to large-scale problems. This versatile numerical method has been extensively used in a diverse range of fields, including fluid mechanics, astrophysics, geosciences and mathematical biology, among many other. The Radial Basis Function-generated Finite Difference Method (RBFFD) is essentially a local approximation of RBF, significantly reducing computational costs, often by several orders of magnitude. Specifically, the cost benefit can be estimated as $N/n$, where $N$ is the total number of nodes—typically ranging from thousands to millions—and $n$ is the number of nearest neighbors considered in the finite difference calculation, usually in the dozens.
As a mesh-free numerical approach of the RBFFD method allows for its application to arbitrary geometries since nodes can be placed freely, being unconstrained by grids or geometry of computational boxes \cite{fornberg2015primer}. Furthermore, high orders of accuracy, ranging from 4th to 8th order, can be easily achieved within the RBF framework. More importantly, it performs exceptionally well for parallelization on distributed memory computers and GPUs. Finally, yet another advantage is that the local RBF node refinement becomes straightforward without the necessity of mesh construction, regardless of the number of dimensions. Despite of many benefits offered by the RBF-based methodology, its use in soft condensed matter so far is limited to vectorial Frank-Oseen-based modeling of solitonic director structures \cite{sohn2018dynamics,durey2020topological}, to the best of our knowledge. At the same time, the more complex tensorial Landau-de Gennes  modeling of orientationally ordered media by utilizing the RBF-based methodology has not been demonstrated, which is the goal of this our work.

In this study, taking advantage of the geometric flexibility of the RBF-based methodology, we show how RBFFD facilitates tensor-based numerical modeling is utilized for uncovering field configurations in confined liquid crystals (LCs) with embedded colloids of complex shapes or inside droplet-confining volumes with nontrivial topologies. The advantages of using RBFs in LC simulations extend to simpler and faster algorithms for adaptive meshing, which is crucial for performing local refinements near the cores of singular defects and geometric features of confining surfaces. The higher accuracy and potential for parallelization enable the modeling of larger LC systems (as well as LC colloidal dispersions, emulsions and other types of complex composite soft matter).
While the application of the RBFFD method in $\bQ$-tensor-based liquid crystal simulations has not been documented in the literature to our knowledge, the RBFFD method has been successfully implemented in various range of fields for solving partial differential equations (PDEs) with flexible geometries. 
These include numerical computation of thermal convection, reaction-diffusion equations, Navier–Stokes equations, and electromagnetics in materials modeling, astrophysics, geophysics, biology, among many others \cite{park1991universal,flyer2016enhancing,piret2012orthogonal,wright2009onset,fornberg2011stabilization,bayona2010rbf,prieto2016rbf,parand2011improved,hillier2014three,sohn2018dynamics,lehto2017radial,lai2008meshless,martin2015seismic}.
Thus, the current simulation framework for LC $\bQ$-tensor modeling can be significantly enhanced by incorporating the RBFFD method. Irregular node sets of the RBF-based techniques are also beneficial to numerical calculations because they do not impose artificial symmetries which introduce undesired numerical artifacts. Therefore, overall, the RBF-based techniques have the potential to improve the performance of numerical simulations of LCs, as we demonstrate below.

\section{Methods}
\label{method}

\begin{figure}
	\includegraphics[width=0.5\textwidth]{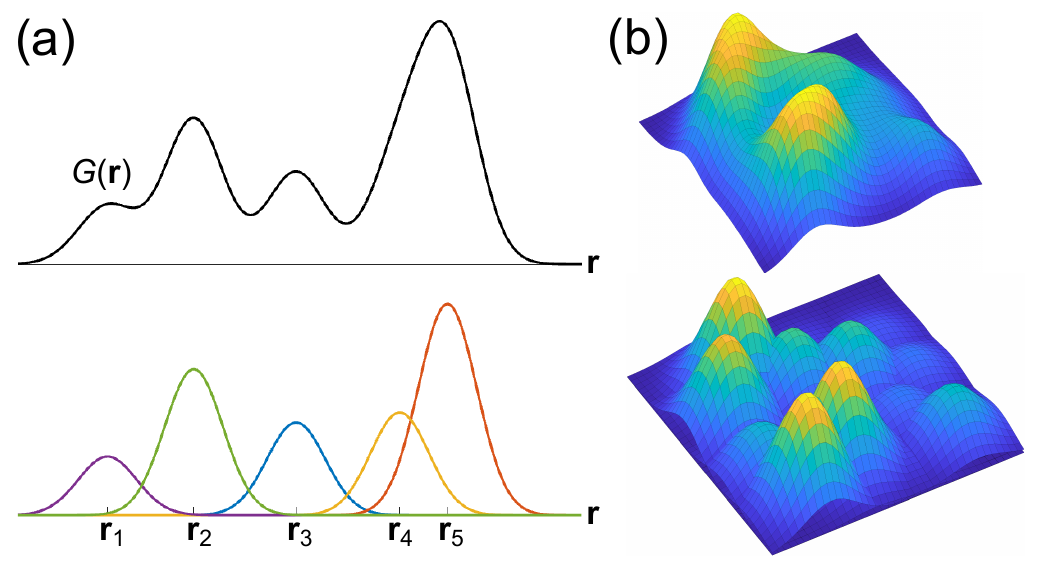}
	\caption[Example superpositions of radial basis functions]{Example superpositions of radial basis functions in one-  (a) and two- (b) dimensional systems.
    }
    \label{rbf_example}
\end{figure}

\subsection{Nematic liquid crystal free energy functional}
Equilibrium structures of LCs were obtained through numerical minimization of liquid crystal energies. First, the free energies are expressed in terms of the Landau-de Gennes expansion of the bulk free energy densities. These includes the thermotropic terms that govern the isotropic-nematic phase transition
\begin{equation}
    f_\rm{thermo} = \frac{A}{2}Q_{ij}Q_{ji}
    + \frac{B}{3} Q_{ij}Q_{jk}Q_{ki}
    + \frac{C}{4} ({Q_{ij}Q_{ji}})^2
\end{equation}
and one-constant approximation of the elastic energy arising from the deformation to the director alignment
\begin{equation}
    f_\rm{elastic} = L \partial_k Q_{ij}\partial_k Q_{ij}
    \label{Qenergy}
\end{equation}
where $\bQ$ is the tensorial order parameter for the moelcular director; $A$, $B$, and $C$ are parameters related to the isotropic-uniaxial phase behavior of CLCs
\cite{gramsbergen1986landau,vissenberg1997generalized,allender2008landau}; $L$ is an elastic constant determining the associate energy costs of spatial distortions \cite{mori1999multidimensional,wu2025topological,lapointe2010electrically}.
Besides, surface boundary conditions at colloidal-molecular interfaces are added to the calculation as conically degenerate surface anchoring described by \cite{zhou2019degenerate}:
\begin{align}
    f_\rm{surf}=W \left ( P_{ik} \tilde{Q}_{kl} P_{lj} - \frac{3}{2}\Seq\cos^2\theta_\rm{e} P_{ij} \right )^2
    \label{LdGsurf}
\end{align}
with $W$ the anchoring strength, $\bf{P}=\bf{v} \otimes \bf{v}$ the surface projection tensor, $\bf{v}$ the surface normal director, and $\tilde{\bf{Q}}=\bQ+\frac{1}{2}\Seq\bf{I}$. Here, $f_\rm{surf}$ is independent of the azimuthal angle of the principal molecular axis with respect to the surface normal vector. If not stated otherwise, planar degenerate anchoring $\theta_\rm{e}=90\dg$ was adopted in our simulations.
Finally, we obtained the total energy by the integration of the domain
\begin{equation}
    F_\rm{total}=\int d\br^3 ~(f_\rm{thermo}+f_\rm{elastic}) + \int d\br^2 ~f_\rm{surface}
    \label{ftotal}
\end{equation}
and equilibrium structures were found based on the gradient descent method
\begin{equation}
    \frac{d\bQ}{dt} \propto -\frac{\delta F_\rm{total}}{\delta \bQ}
    \label{dfdQ}
\end{equation}
with adaptive Runge-Kutta integration (ARK23) over the imaginary timescale $t$. Fast inertia relaxation engine (FIRE) was incorporated as well \cite{sussman2019fast}. A typical numerical iteration of energy minimization contains one ARK step and ten FIRE steps.

The initial configurations were unidirectional with the molecular alignment along $\nv_0$ for LC simulations involving colloidal spheres and handlebodies (\figs{handle}a, \ref{sphere}, and \ref{sphere_regular}). 
In contrast, the local director field conditions represented within the nematic volume were initially oriented perfectly perpendicular to the center along the circumferences for donut-shaped LC droplets (\fig{droplet}).
For a typical computation of LC structure, the following parameter values were used \cite{ravnik2009landau,mucci2016landau,wu2025topological,lapointe2010electrically}:  $A=-1.72 \times 10^5 \Jm{-3}$, $B=-2.12 \times 10^6 \Jm{-3}$, $C=1.73 \times 10^6 \Jm{-3}$, $L=20 \pN$, and $W$ ranges from $2\times10^{-4}$ to $10^{-2}\Jm{-2}$. The equilibrium $S_\rm{eq}\approx 0.533$.

\subsection{Radial basis function and finite difference coefficient}
Similar to the Fourier transformation, we represent position-dependent values, such as the components of the $\bQ$ tensor, as a linear combination of $n$ independent functions \cite{fornberg2011stabilization,fornberg2015solving,sohn2018dynamics,flyer2016enhancing}
\begin{equation}
    G(\br) \approx \sum_{i=1}^{n} g_i \phi_i(\br)=\sum_{i=1}^{n} g_i \phi(\br-\br_i)
    \label{rbf}
\end{equation}
where $\phi_i$ are the radial basis functions and $g_i$ are scalar coefficients (\fig{rbf_example}). In fact, pseudospectral methods, including Fourier transformation, can be viewed as the limit of RBF method \cite{fornberg2015primer}, whereas in conventional computations on regular grids, the Dirac delta function is chosen at each grid point $\phi_i(\br-\br_i)=\delta(\br-\br_i)$.
For RBFFD method, common choices of the radial basis functions are those exhibiting azimuthal symmetry $\phi(\br-\br_i)=\phi(|\br-\br_i|)$. Examples include multiquadric, inverse quadric, and polyharmonic spline functions. These functions are spatially localized, in contrast to the sinusoidal waves used in Fourier transformation, allowing for finite difference approximations to be performed using local neighboring values (see below for details). In this work, we select the Gaussian function $\phi(\br-\br_i)=e^{-\epsilon|\br-\br_i|^2}$ due to its infinite smoothness (spatial derivatives of any order are smooth functions) and the ease of calculating derivatives. We also adopted a shape parameter $\epsilon=0.3-0.5$ when the coordinate $\br$ is rescaled by the average node separation distance (typically tens of nanometers).
The center positions of these radial basis functions will be referred to as nodes or node positions for the remainder of the article.

The finite difference approximation for any linear operator $D$, including differential operators that essential in $\bQ$ tensor simulation, is performed via
\begin{align}
    \left.D[G(\br)]\right|_{\br=\br_\rm{c}}
    = \left.\sum_{i=1}^{n} g_i D[\phi(\br-\br_i)]\right|_{\br=\br_\rm{c}} \nonumber \\ 
    \approx \sum_{j=1}^{n}d_jG(\br_j) 
    = \sum_{i=1}^{n} \sum_{j=1}^{n} g_i d_j \phi(\br_j-\br_i)
    \label{fd}
\end{align}
where $\br_\rm{c}$ is the position of evaluation and $d_{j}$ are finite difference coefficients. $d_{j}$ are obtained by asserting \eq{fd} to be exact for all RBF interpolants $G(\br) \equiv \phi(\br-\br_i)$. In other words, we solve $n$ linear equations represented by \cite{fornberg2015solving,sohn2018dynamics}
\begin{align}
    \begin{bmatrix}
    \phi(\br_1-\br_1) & \phi(\br_2-\br_1) & ... &  \phi(\br_n-\br_1) \\
    \phi(\br_1-\br_2) & \phi(\br_2-\br_2) & ... &  \phi(\br_n-\br_2) \\
    \phi(\br_1-\br_3) & \phi(\br_2-\br_3) & ... &  \phi(\br_n-\br_3)\\
    \vdots & \vdots & \ddots & \vdots \\
    \phi(\br_1-\br_n) & \phi(\br_2-\br_n) & ... &  \phi(\br_n-\br_n)\\
    \end{bmatrix}
    \begin{bmatrix}
    d_1 \\ d_2 \\ d_3 \\ \vdots \\ d_n
    \end{bmatrix}
    \nonumber \\
    \equiv 
    A
    \begin{bmatrix}
    d_1 \\ d_2 \\ d_3 \\ \vdots \\ d_n
    \end{bmatrix} 
    =
    \left.
    \begin{bmatrix}
    D[\phi(\br-\br_1)] \\ 
    D[\phi(\br-\br_2)] \\
    D[\phi(\br-\br_3)] \\
    \vdots \\
    D[\phi(\br-\br_n)]
    \end{bmatrix}
    \right|_{\br=\br_\rm{c}}
    \label{fdcoef}
\end{align} 
for the $d_{j}$ coefficients. 
The $A$-matrix and the ensuing derivative coefficients are determined by the choice of the radial basis function $\phi$ and the linear operator $D$ while independent of the actual function values $g_i$.
The nonsingularity of the $A$-matrix is guaranteed regardless of how nodes are arranged across multiple dimensions \cite{fornberg2015solving}, which circumvents the singularity problems commonly encountered in pseudo-spectral methods when applied to node sets in two or more dimensions.
To enhance the performance and increase the accuracy of the numerical computation, we incorporated supplementary polynomials up to the second degree into the RBF basis in \eq{fdcoef} \cite{sohn2018dynamics,bayona2017role}
\begin{align}
    \begin{bmatrix}
    \begin{array}{c|c}
    A & 
    \begin{matrix}
        1 & x_1 & x_1^2 & y_1 & y_1^2 & z_1 & z_1^2 \\
        1 & x_2 & x_2^2 & y_2 & y_2^2 & z_2 & z_2^2 \\
        \vdots & \vdots & \vdots & \vdots & \vdots & \vdots & \vdots \\
        1 & x_n & x_n^2 & y_n & y_n^2 & z_n & z_n^2
    \end{matrix}
    \\
    \hline
    \begin{matrix}
        1 & \cdots & 1 \\
        x_1 & \cdots & x_n \\
        x_1^2 & \cdots & x_n^2 \\
        y_1 & \cdots & y_n\\
        y_1^2 & \cdots & y_n^2 \\
        z_1 & \cdots & z_n \\
        z_1^2 & \cdots & z_n^2
    \end{matrix}
    & 0
    \end{array}
    \end{bmatrix}
    \begin{bmatrix}
    d_1 \\ d_2 \\ d_3 \\ \vdots \\ d_n \\ d_{n+1} \\ \vdots \\ d_{n+7}
    \end{bmatrix}
    \nonumber \\
    =
    \left.
    \begin{bmatrix}
    D[\phi(\br-\br_1)] \\ 
    D[\phi(\br-\br_2)] \\
    D[\phi(\br-\br_3)] \\
    \vdots \\
    D[\phi(\br-\br_n)] \\
    D1 \\
    Dx \\
    Dx^2 \\
    Dy \\
    Dy^2 \\
    Dz \\
    Dz^2
    \end{bmatrix}
    \right|_{\br=\br_\rm{c}}
    \label{fdcoef2}
\end{align}
where $\br_i=\{x_i,y_i,z_i\}$.
For each node point $\br_\rm{c}$, we considered 24 nearest neighbors ($n=24$) when solving \eq{fdcoef2} and the unique set of the $d_j$ coefficients were used as long as all the 25 node positions (24 neighbors and self) remained unchanged. In the adaptive node refinement detailed below, however, any change to these node positions requires re-evaluation of \eq{fdcoef2} and updates to the $d_j$ coefficients.

\subsection{Adaptive node set}
Adaptive node distribution in a liquid crystal numerical volume is essential for refining details near strong elastic distortions, such as defects, while maintaining reasonable computational efficiency. Adaptive grid sets have been extensively studied and implemented in the finite element approach for Landau–de Gennes modeling of nematic liquid crystals \cite{amoddeo2024moving,tasinkevych2012liquid}. In this work, we aimed to optimize the node set during energy relaxation, as described in \eq{dfdQ}, by establishing an ideal node separation distance that is proportional to the greatest second derivatives of the $\bQ$ tensor
\begin{equation}
    r_0 \propto (\partial_i\partial_j Q_{ab})^{-1/2}
\end{equation}
We then calculated electrostatic-like repulsive forces between nodes, with each node carrying a ``charge'' equal to the corresponding ideal node separation distance $r_0$. By incorporating node movement during the minimization of the free energy \eq{ftotal}, the nodes adaptively relocate to regions of the liquid crystal with stronger distortions in the order parameter $\bQ$. As a result, we obtained node spacings ranging from 1.5 to 7 nm in our findings. Although the node positions were initially set as randomly placed points, we employed a similar node refinement method based on electrostatic-like repulsions to effectively achieve a uniformly distributed node set ahead of the $\bQ$ energy minimization. In our study, $N=8\times 10^5$ to $1.5\times 10^6$ node points were typically used for each simulation.

\subsection{Experimental Sample Preparation and Imaging Methods}
    In addition to comparisons of the RBFFD-based results with ones obtained by other numerical methods, we also compared them to experimental findings, especially in cases of LC colloids and droplets with nontrivial topology, like the handlebodies of different genus values. Microfabrication of colloidal handlebodies followed a procedure where a thin SiO\textsubscript{2} layer was deposited on a silicon wafer using plasma-enhanced chemical vapor deposition  \cite{liu2013nematic}. Then, a 1.5-micrometer-thick layer of SU-8 photoresist (Microchem) was spin-coated onto the SiO\textsubscript{2} layer and the pattern of rings was defined by using a standard photolithographic technique. The SiO\textsubscript{2} layer was then wet-etched in a buffered solution (HF:NH\textsubscript{4}F:H\textsubscript{2}O = 3:6:10 by weight) \cite{liu2013nematic}.  The handlebody particles made of SU-8 polymer were released into the solution in this process. After washing in deionized water and re-dispersing these particles in E7 or 5CB (from EM Chemicals), the nematic dispersion was infiltrated into glass cells comprised of glass plates separated by glass spacers defining the cell gap\cite{liu2013nematic}.  Substrates of these glass cells were coated with polyimide PI2555 (HD Microsystems) for in-plane alignment of the far-field director defined by rubbing. 
    To create handlebody-shaped droplets of LC inside of the polymer matrix, we first fabricated handlebody-shaped silica microstructures using photolithography \cite{liu2013nematic, campbell2014topological}.  A droplet of Norlin Optical Adhesive (NOA-63) was then pressed against the substrate containing these silica microstructures. Then, the desired micrometer-sized polymer confinement structures were obtained using replica molding, where the UV-curable glue NOA-63 was cured by using the OmniCure S2000 illumination system (pur hased from Lumen Dynamics) by high-intensity 350-400nm UV radiation \cite{campbell2014topological}. After curing for about 30 seconds, the polymer sheet was pealed off while having the desired surface topography on one of its sides. We also fabricated flat thin polymer microfilms by curing NOA63 between two flat substrates at the same illumination\cite{campbell2014topological}. To create polymer films containing nematic drops, we infiltrated room-temperature nematic LC mixtures E7 or E31 (from EM Chemicals) into a gap between a flat polymer sheet and the film with microstructures and pressed the two together while using additional UV-curing\cite{campbell2014topological}.  This procedure yields polymer films containing LC drops with different genus of confining surfaces.  The details of such sample preparation procedures are described elsewhere. 
    The experimental structural analyses were based mainly on the three-photon excitation fluorescence polarizing microscopy (3PEF-PM) based imaging implemented using an inverted microscope IX 81 (Olympus), a tunable Ti-Sapphire oscillator (680‑1,080 nm, Coherent) emitting 140 fs pulses at a repetition rate of 80 MHz and an oil-immersion 100x objective with a numerical aperture of 1.4  \cite{liu2013nematic, campbell2014topological}.   The detected 3PEF-PM intensity exhibited a strong dependence on the angle between the local orientation of the LC director and the linear polarization of excitation light, allowing reconstruction of structures inside the handlebody shaped droplets and around handlebody shaped particles  \cite{liu2013nematic, campbell2014topological}. We also utilized a conventional polarizing optical microscopy (POM), with and without additional phase retadation plates inserted into the optical path of POM. Both 3PEF-PM and POM images could be then directly compared to their computer-simulated counterparts. The numerical methods for computer-simulating  POM and 3PEF-PM images based on energy minimizing structures obtained using RBFFD or other techniques are described in detail elsewhere \cite{liu2013nematic, campbell2014topological}.

\section{RBFFD calculations for nontrivial geometries}
\begin{figure*}
    \centering
	\includegraphics[width=1\textwidth]{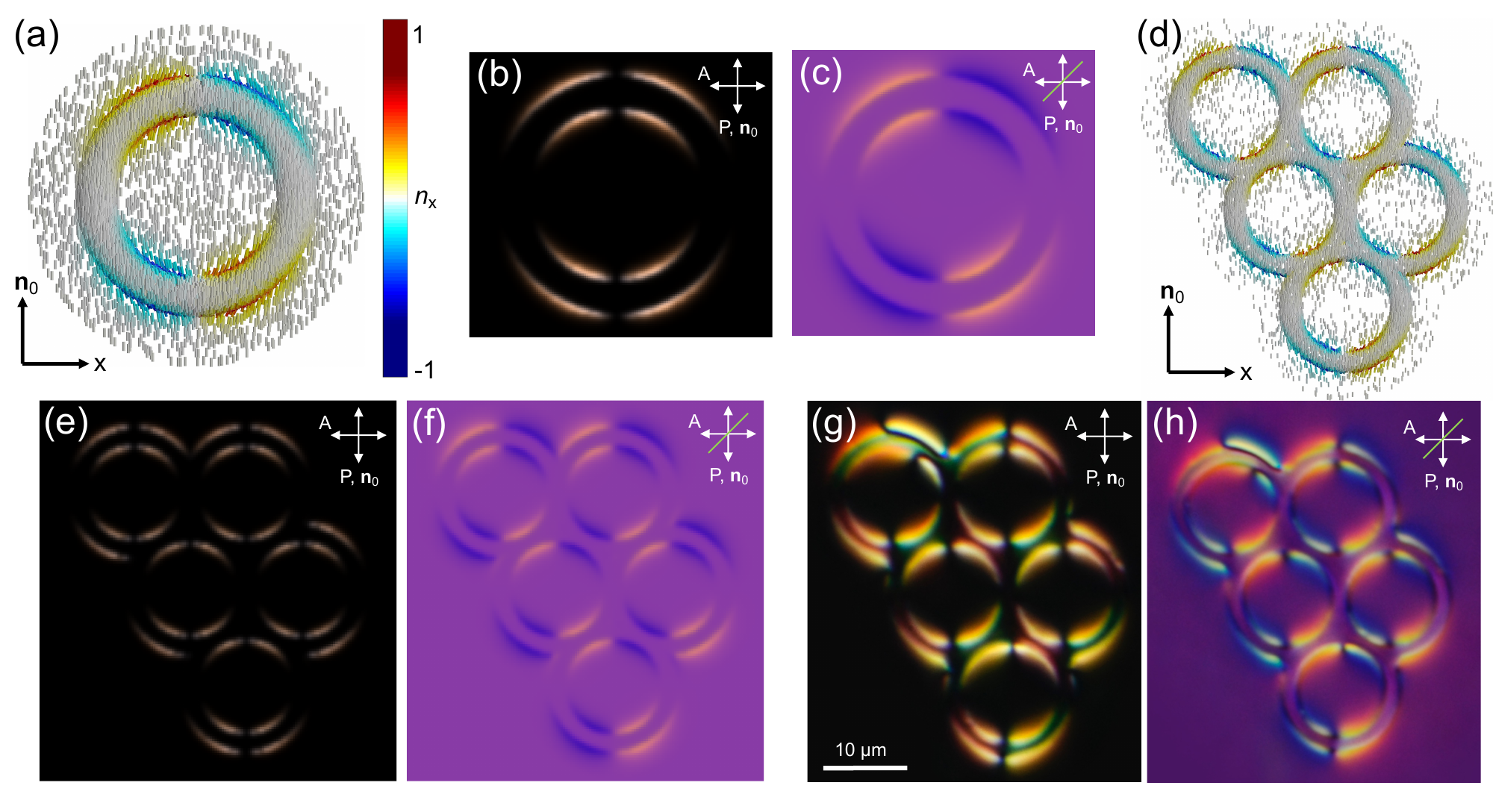}
	\caption[Radial-basis-function simulation of nematic LCs with handlebody geometries]{Radial-basis-function simulation of nematic LCs with handlebody geometries.
    (a) Simulation of a nematic LC initially aligned along $\nv_0$ and then perturbed by a torus-shaped particle with planar degenerate surface anchoring $\theta_\rm{e}=90\dg$. Nematic director field is rendered using cylinder rods, colored by their $x$ projection. 
    (b,c) THe corresponding polarizing optical microscopy (POM) simulations with (b) our without (c) a half-wave plate. 
    (d-f) Simulations with a handlebody of genus-5 with planar degenerate boundary conditions, visualized using cylinders (d) and POMs (e,f).
    (g,h) The corresponding experimental POM micrographs of nematic LCs near the genus-5 surfaces.
    Polarizers (P) and anayzers (A) are marked in each optical images and the green lines at 45\dg indicate the optical axis of half-wave plates when used.
    }
    \label{handle}
\end{figure*}

\begin{figure*}
    \centering
	\includegraphics[width=1\textwidth]{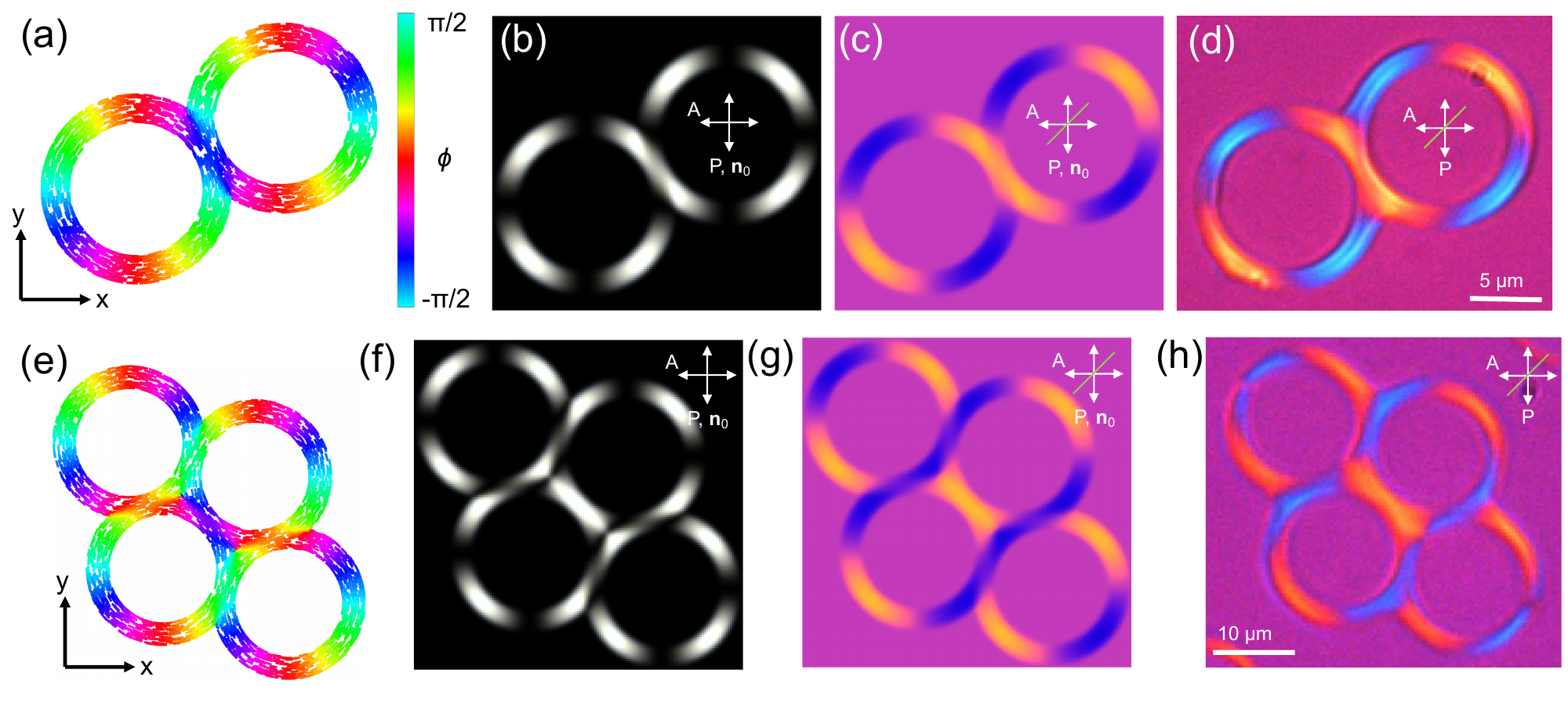}
	\caption[Radial-basis-function simulation of nematic LCs droplets confined in the interior of genus 2 and 4 surfaces]{Radial-basis-function simulation of nematic LCs droplets confined in genus 2 and 4 surfaces.
    (a) RBFFD calculation of a nematic LC confined to the volume of a genus-2 handlebody with the LC-channel interfaces have planar surface anchoring $\theta_\rm{e}=90\dg$. Inset shows the colorscheme for the $x$-$y$ azimuthal angle $\phi$ of the nematic director. 
    (b,c) The corresponding POM simulations with (b) or without (c) a half-wave plate.
    (d) The experimental realization of the nematic LC droplet within the figure-8-shaped geometry imaged using POM.
    (e-h) Similarly presented simulation (e-g) and experiment (h) of LC in tubes with genus-4 surface.
    Polarizers (P) and anayzers (A) are marked in each optical images and the green lines at 45\dg indicate the optical axis of half-wave plates if used.
    }
    \label{droplet}
\end{figure*}

RBFFD benefits significantly from the flexibility of node placement. We therefore conducted calculations on a nematic LC surrounding a handlebody-shaped geometry with planar boundary conditions (\fig{handle}).
In \fig{handle}a, we can observe how the planar degenerate anchoring forces perturb the liquid crystal director field both inside and outside the ring-shaped colloidal particle, inducing molecular director distortions that are clearly visualized in the simulated polarizing optical microscopy (POM) images (\fig{handle}b,c). We also would like to point out that regions with stronger elastic distortions were resolved with a higher density of nodes, as illustrated in \fig{handle}a, as a result of the adaptive node refinement.
Moreover, the meshless node set in the RBFFD calculation allowed us to easily expand the size of our numerical simulations, enabling us to model a liquid crystal system with a particle shaped like a genus-5 handlebody (\fig{handle}d-f). This calculation was accelerated by initializing the $\bQ$ condition as five translated copies of the previously simulated result for a single particle, which required minimal effort within the RBF framework. Good agreements with experimental results were observed based on the POM images (\fig{handle}e-h).

Additionally, we observed that the total surface defect charges comply with topological constraints. According to a generalized hairy ball theorem, a tangent vector field cannot be non-vanishing and continuous everywhere on a closed surface with a nonzero Euler characteristic $\chi=2-2g$, where $g$ is the genus of the surface.
Thus, the handlebody colloids with tangential boundaries, as presented here, are always accompanied by topological defects with a total winding number equal to $\chi=2-2g$ on the particle surfaces \cite{senyuk2013topological,senyuk2015three}, provided that the planar anchoring is sufficiently strong. In our simulation of a genus 1 surface (\fig{handle}a), we counted two $+1$ and two $-1$ surface defects, leading to a total winding number of 0. For the genus 5 handlebody in \fig{handle}d, we found a total winding number of $3(+1)+11(-1)=-8$, consistent with the principles of algebraic topology. Note that each integer defect split into two half-integer defects with a small separation distance in our simulations due to their lower elastic energies. Here we count each pair as one integer defect for convenience.

To demonstrate the flexibility of RBFFD calculations, we present simulations of liquid crystal droplets within circular channels (\fig{droplet}). The liquid crystals were placed inside these tubes, with the inner surfaces exhibiting planar degenerate anchoring ($\theta_\rm{e}=90\dg$). With an initial configuration of the molecular director $\nv$ aligned along the tube axes, singular defects were identified at the joints of the LC-filled pipelines.
\fig{droplet}a-c, for example, show two defects (each with a topological charge of $-1$), which are consistent with both the experimental observations (\fig{droplet}d) and the theorem mentioned earlier ($2-2g=-2=2(-1)$). We also present a similar but larger example using nematic LC filled into four interconnected circles (\fig{droplet}e-g), where twelve $-1/2$ defects surface defects were observed and a total winding of $-6$ was found, as expected. In this case, the positions of the defects were slightly different from those in the corresponding experimental POM images shown in \fig{droplet}h, likely due to the intricate surface geometry.

\section{RBFFD-based analysis of elastic colloidal multipoles}

\begin{figure*}
    \centering
	\includegraphics[width=1\textwidth]{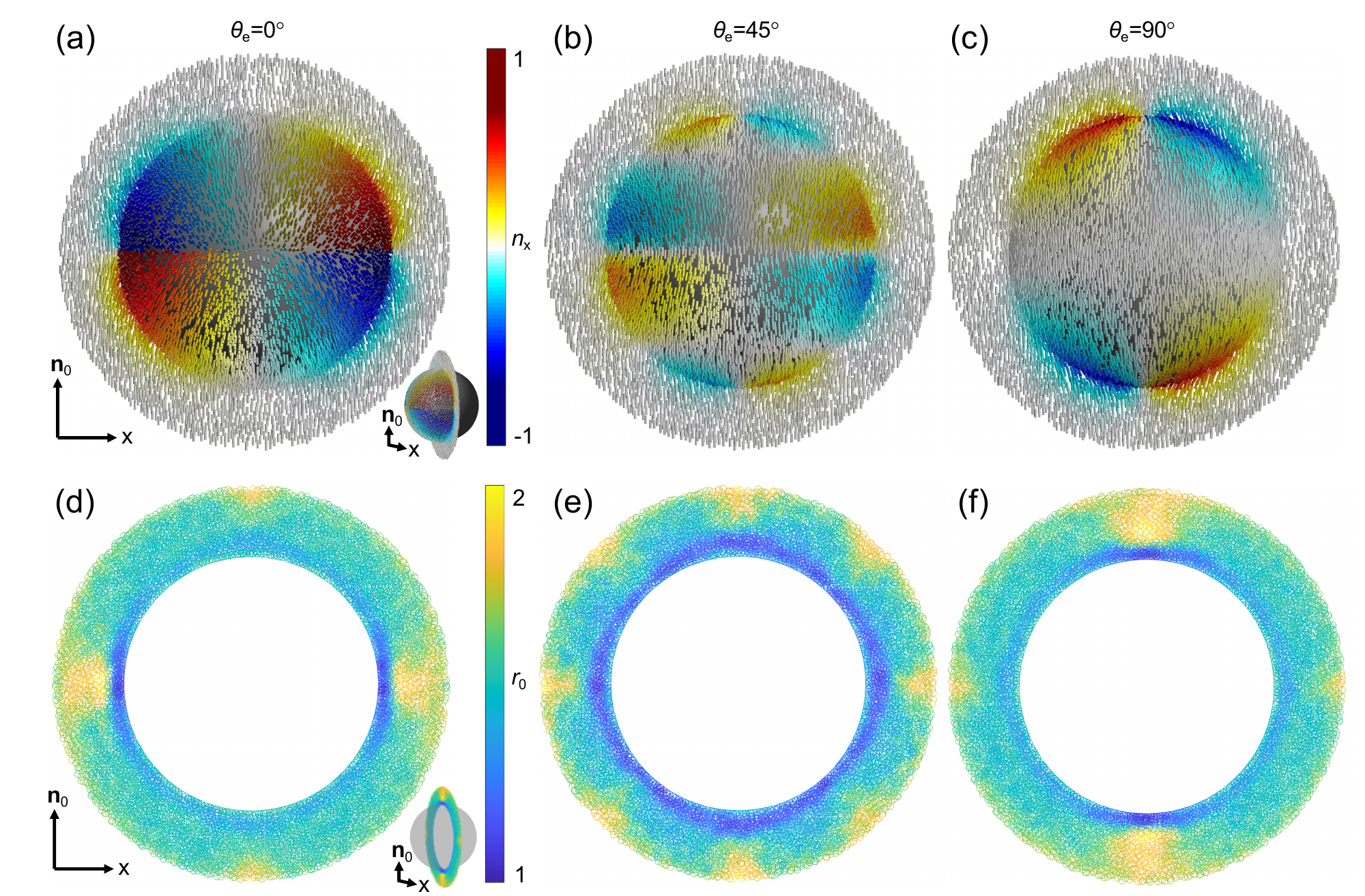}
	\caption[Radial-basis-function-based numerical simulation of LC with colloidal spheres]{Radial-basis-function-based numerical simulation of LC with colloidal spheres of different anchoring conditions.
    (a-c) Computed nematic director $\nv$ field around a sphere immersed in a uniformly aligned nematic director field $\nv_0$, showing elastic quadrupoles for homeotropic $\theta_\rm{e}=0\dg$ (a) and degenerate planar $\theta_\rm{e}=90\dg$ (c) anchoring, and a hexadecapole for degenerate conic anchoring $\theta_\rm{e}=45\dg$ (b). Equilibrium surface anchoring angle $\theta_\rm{e}$ between molecular director $\nv$ and surface normal vector are marked for each case. Local orientations of the director field are visualized on the spherical surfaces as well as on $n_0$-$x$ cross-sections, as shown by the inset in (a). Director field local orientations are represented by cylinders colored according to their $x$ component $n_x$ with the colormap in (a). Surface anchoring coefficient $W=2\times10^{-4}\Jm{-2}$ and sphere radius $r_\rm{c}=5\um$.
    (d-f) The corresponding local average of node separation distance on the $n_0$-$x$ cross-sections for each simulation with different $\theta_\rm{e}$ colored by rescaled node separation distances $r_0$.
    }
    \label{sphere}
\end{figure*}

\begin{figure*}
    \centering
	\includegraphics[width=1\textwidth]{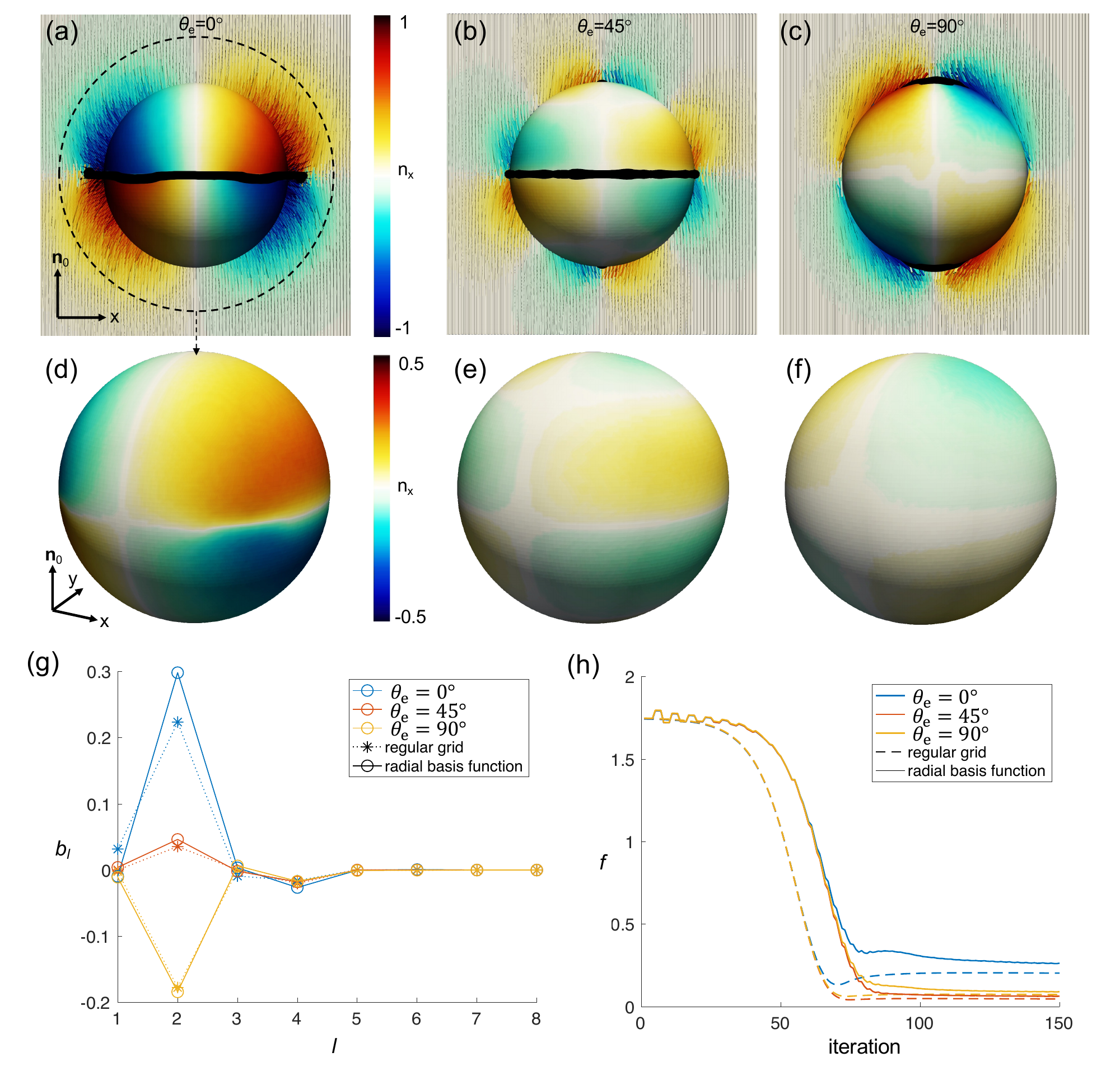}
	\caption[Nematic LC simulations based on regular grid points and comparison to RBFFD simulations]{Nematic LC simulations based on regular grid points and comparison to RBFFD simulations.
    (a-c) Numerical simulation in regular grids of LC surrounding a spherical particle. Particle surfaces and director fields are colored by the $x$ projection of $\nv$, and blackened regions of the surface correspond to singular defects. Far-field LC director orientation $\nv_0$ and $x$ axis are labeled in (a). Surface anchoring coefficient $W=2\times10^{-4}\Jm{-2}$ and sphere radius $r_\rm{c}=5\um$.
    (d-f) Corresponding particle-induced LC elastic distortions at a radius $r = 1.5r_\rm{c}$, whose distance is marked in (a).
    (g) Elastic multipole moment $b_l$ calculated for LC systems shown above. Dashed lines are results obtained from regular grid sets whereas solid lines with circles are from radial-basis-function finite difference (RBFFD) calculations (\fig{sphere}).
    (h) Rescaled LC free energies $f$ over number of iterations during numerical energy relaxation.
    All numerical simulations are performed under one-constant approximation (see \sect{method}).
    }
    \label{sphere_regular}
\end{figure*}

We represent the elastic multipoles formed around spherical colloidal particles as another demonstration of RBFFD calculations (\fig{sphere}). The elastic distortions in the molecular director field $\nv$ are induced by various types of surface anchoring, including perpendicular boundary conditions ($\theta_\rm{e}=0\dg$, \fig{sphere}a,d), conically degenerate ($\theta_\rm{e}=45\dg$, \fig{sphere}b,e), and planar degenerate ($\theta_\rm{e}=90\dg$, \fig{sphere}c,f). We clearly observe that the shapes of the elastic deformations resemble quadrupoles for $\theta_\rm{e}=0\dg$ and $90\dg$, and a hexadecapole for $45\dg$ \cite{zhou2019degenerate}.
Furthermore, \fig{sphere}d-f illustrate the local node separation distances, which is calculated by the average distance to the 24 nearest neighbors for each node point. It is evident that node concentrations increase near defects close to the spherical surfaces, overlapping with the regions with strong director distortions (dark blue and red in \fig{sphere}a-c). For example, \fig{sphere}d shows two defects on the left and right sides of the sphere, corresponding to the intersections of the Saturn defect ring with the $\nv_0$-$x$ plane of visualization \cite{zhou2019degenerate}. Similarly, \fig{sphere}f depicts two boojums at the north and south poles of the spherical surface, where dense clusters of nodes accumulate near the surface defects.

Interestingly, the symmetries of the liquid crystal systems are also clearly manifested in regions of sparser nodes or larger node separations, apart from the defect areas. The relatively lower node concentrations indicate minimal elastic deformations and mark the volumes where the molecular director $\nv$ remains parallel to the background alignment $\nv_0$. Specifically, \fig{sphere}d and f both display four yellow regions on the $\nv_0$-$x$ cross-section, aligned symmetrically and representing quadrupoles. In contrast, the eight uniformly distributed sparse regions around the circumference in \fig{sphere}e clearly indicate a hexadecapole.
Since the node sets for these RBFFD simulations were initialized using random positioning with frequent updates to their locations (\sect{method}), the symmetries observed around each spherical particle arise solely from the elastic deformations and defect distributions, in contrast to simulations conducted in regularly meshed boxes where undesired artifacts may arise due to the grid geometry.

For benchmarking purposes, the elastic multipoles simulated using cubic meshes are presented in \fig{sphere_regular}a-f, closely matching the previously presented results from RBFFD calculations \fig{sphere} by exhibiting similar director distortions, defect positions, and broken symmetries. Besides, we evaluate the multipole moments in order to further quantify such elastic distortions of nematic director. 
It is known that under the one-constant approximation of the elasticities, these elastic distortions can be effectively represented by multipole expansions, as the elastic energy reduces to a simple Laplace equation \cite{senyuk2019high,pergamenshchik2014elastic}. Despite the additional thermotropic energies included in this study, we can still approximate the obtained structures using multipole expansions of the liquid crystal director field $\nv$. With the approach detailed in Ref.~\cite{senyuk2019high}, we selected a radius where the director deviation is small $n_x \ll 1$ and computed the multipole moments $b_l$ by decomposing $n_x(\br)$ into spherical harmonics. The results are plotted in \fig{sphere_regular}g for both RBFFD and conventional finite difference methods, showing a good agreement between the two approaches. Since the elastic multipoles are dominantly quadrupolar ($l=2$) for $\theta_e=0\dg$ and $90\dg$ \cite{lev2002symmetry,lapointe2009shape} and hexadecapolar ($l=4$) for other oblique angles \cite{senyuk2016hexadecapolar}, we conclude that our RBFFD simulations of the nematic LC system accurately capture the expected features.

Also, we observed similar energy conversion during the structure relaxation of the nematic LC as shown in \fig{sphere_regular}h. The rescaled energy density $f$ is the total free energy of the LC-particle system divided by the number of active LC points. Starting at the same (background) value, the free energies decreased over numerical iterations and reached a steady value, based on which we determined the equilibrium state. In our implementation of fast inertia relaxation engine (\sect{method}), the step size in time (\eq{dfdQ}) can smoothly grow over iterations to accelerate the energy conversion. On the other hand, the adaptive node optimization in the RBFFD approach where nodal points clustered around defects slows down the energy minimization by limiting such growth of the time step size, which explains the discontinuous jumps in the energies and their slower conversion to equibrium values (\fig{sphere_regular}h). We also attributed the slightly higher equilibrium (steady) energies in the RBFFD calculations to more refined node sets and more accurate representations of LC molecular alignment fields around defects as compared to the calculations based on regular grids \fig{sphere_regular}h. 

\section{Discussion}
Our RBFFD implementation in nematic liquid crystal simulations provides an accurate and flexible alternative, maintaining both low computational cost and high efficiency.
Among others, the greatest advantage our RBFFD calculations offer comes from the flexible, adaptive node positions with fewer node points.
For example, the numerical simulations on a $120^3$ cubic grids in \fig{sphere_regular} require more than twice as many data points to represent the same geometry compared to the RBFFD approach (800,000 used for those in \fig{sphere}). The inclusion of the inactive volume inside spherical particles primarily contribute this, while no redundant nodal points are needed in RBFFD.
With an adaptive node set, we also reached higher resolution near singular defects while sparser node swts were used elsewhere without implementing multigrid method explicitly.

Furthermore, we found that RBFFD offers approximately 50\% less computation time in our test with the same number of iterations, provided a fixed node set is used without adaptive node optimization. This improvement of computation speed from RBFFD method is primarily due to the reduced number of nodes and volume needed, which is half of those used in regular FD.
Moreover, the computational cost associated with node optimization comes from the calculation of the differential coefficients in \eq{fdcoef}, which necessitates evaluations of the matrix inversion by $\mathcal{O}(N)$ times. 
Calculation for updating node positions, however, benefit substantially from GPU and parallel computing, since the differential coefficient calculation for each node is independent of each other and can be computed simultaneously.
Using our home-built MATLAB code run on a single CPU, we found that the calculation of node optimization \eq{fdcoef} takes roughly $30\times$ more computation time compared to director optimization \eq{dfdQ}, which would become much less costly depending on how many processor/threads implemented. In our test with a size of ten thousand nodes, we found an overall 33\% speed improvement using three parallel processes.
We employed three iterations of energy optimizations before each node optimization in the presented results, hence the irregular meander in the energy profile \fig{sphere_regular}h.
Further optimization on the partition between director and node position calculations is needed, though, to efficiently utilize computational power while still benefiting from adaptive nodes for higher accuracy.

In terms of computer memory management, though, the memory usage for RBFFD calculations may be higher than that of conventional methods, due to the storage of the differential coefficients $d$, which requires a size of $N\times n$ ($N$ is the number of active node and $n$ is the number of neighbors considered in finite difference) times a numerical scalar value. In contrast, the cubic grid has all finite difference coefficients explicitly calculated, thereby avoiding this cost.
Nevertheless, with parallel computing, the usage of RBFFD method has significantly accelerated our LC simulation while maintaining high order of accuracy.
However, we do not aim to conduct a robust examination of the calculation accuracy of our RBFFD implementation at this stage. Instead, we present the observed qualitative and quantitative agreement in the benchmarking results summarized in \fig{sphere_regular}g,h, leaving a more detailed investigation for future works. 

\section{Conclusion}
Our Landau-de Gennes numerical simulation of nematic liquid crystals based on radial basis functions offers several advantages over traditional calculations conducted on cubic grids, including greater flexibility in geometry and improved efficiency. Through our demonstrations using LC-colloid and LC emulsion systems with surfaces of varying genus, we have concluded that our calculations of the liquid crystal equilibrium structures accurately capture all expected features across the studied systems.
Additionally, we implemented an adaptive refinement of RBFFD nodes, enhancing our ability to resolve singular defects in nematic liquid crystals. The potential and limits of the adaptive scheme have yet to be tested in scenarios involving highly anisotropic geometries, which will be the focus of our future studies. With further optimizations and parallelizations of the numerical computation schemes, superior performance, in terms of both efficiency and accuracy, can be achieved. While the tensorial modeling of liquid crystals is often limited to relatively small systems due to high computational costs, the demonstrated here RBFs-based implementation of such modeling opens doors to computational studies of larger systems while resolving fine details with the help of adaptive mesh refinements. Furthermore, our RBF-based implementation of modeling may allow for combining tensorial and vectorial approaches, as well as the use of such different techniques adaptively for different sub-volumes of the overall studied systems (e.g., this could be useful in cases of co-existence of singular defects and topological solitonic structures, as in the case of recently studied mobiusons \cite{zhao2023liquid}).

\section*{Acknowledgments}
I.I.S. and J.-S.W. acknowledge the support by the U.S. Department of Energy, Office of Basic Energy Sciences, Division of Materials Sciences and Engineering, under contract DE-SC0019293 with the University of Colorado at Boulder.  I.I.S. acknowledges the support of the International Institute for Sustainability with Knotted Chiral Meta Matter (WPI-SKCM2), an International Institute of Japan's World Premier Initiative during his asabbatical stay. We have greatly benefited from discussions with M. Dijkstra, B. Fornberg, M. Tasinkevych, M. Campbell, S. Zumer, B. Senyuk, H. Zhao, T. Lee and P. Ackerman, as well as we are grateful for the technical assistance of M. Campbell, Q. Liu and B. Senyuk. 
\bibliographystyle{unsrt}
\bibliography{refs}

\end{document}